\documentclass[useAMS,usenatbib]{mnras}
\usepackage{graphicx}
\usepackage{array,longtable}
\usepackage{natbib}
\usepackage{float}
\usepackage[T1]{fontenc}
\usepackage{ae}
\usepackage{aecompl}
\usepackage{amsmath, amssymb, amsfonts}
\usepackage{multicol}
\usepackage{url}
\def\actaa{Acta. Astronom.} %
\def\aj{AJ}%
\def\apj{ApJ}%
\def\apjl{ApJ}%
\def\apjs{ApJS}%
\def\ao{Appl.~Opt.}%
\def\aap{A\&A}%
\def\aaps{A\&AS}%
\def\mnras{MNRAS}%
\def\na{NewA}%
\def\pasp{PASP}%

\title[Extinction Curve Variations in $VIJK_{s}$]{\vspace{-0.7cm} Interstellar Extinction Curve Variations Toward the Inner Milky Way: A Challenge to Observational Cosmology\vspace{-0.6cm}}
\author[Nataf et al.]{David M. Nataf$^{1,2}$\thanks{Email: david.nataf@anu.edu.au},  
Oscar A. Gonzalez$^3$,  
Luca Casagrande$^{1,2}$, 
Gail Zasowski$^{4,5,2}$,
\newauthor{Christopher Wegg$^{6}$,
Christian Wolf$^{1}$,
Andrea Kunder$^{7}$,
Javier Alonso-Garcia$^{8,9}$,  }
\newauthor {Dante Minniti$^{9,10,11}$,  
Marina Rejkuba$^{12,13}$,
Roberto K. Saito$^{14}$,
Elena Valenti$^{12}$,  }
\newauthor {Manuela Zoccali$^{15,9}$ ,
Rados{\l}aw Poleski$^{16,17}$, 
Grzegorz Pietrzy{\'n}ski$^{16,18}$, 
Jan Skowron$^{16}$,    } 
\newauthor {Igor Soszy{\'n}ski$^{16}$,  
Micha{\l} K. Szyma{\'n}ski$^{16}$, 
Andrzej Udalski$^{16}$, 
Krzysztof Ulaczyk$^{16,19}$, }
\newauthor {{\L}ukasz Wyrzykowski$^{16}$}
\vspace*{5pt}\\
$^{1}$Research School of Astronomy and Astrophysics, Australian National University, Canberra, ACT 2611, Australia \\ 
$^{2}$Kavli Institute for Theoretical Physics, University of California, Santa Barbara, CA 93106 \\ 
$^{3}$European Southern Observatory, 19001 Casilla Santiago 19, Chile \\ 
$^{4}$NSF Astronomy and Astrophysics Postdoctoral Fellow \\
$^{5}$Department of Physics \& Astronomy, Johns Hopkins University, Baltimore, MD, 21218, USA \\
$^{6}$Max-Planck-Institut f\"ur  Extraterrestrische Physik, Giessenbachstrasse, 85748 Garching, Germany \\
$^{7}$Leibniz-Institut f\"ur Astrophysik Potsdam (AIP), An der Sternwarte 16, D-14482 Potsdam, Germany \\
$^{8}$Unidad de Astronomia, Facultad Cs. Basicas, Universidad de Antofagasta, \\
Avda. U. de Antofagasta 02800, Antofagasta, Chile \\
$^{9}$Millenium Institute of Astrophysics, Av. Vicu\~na Mackenna 4680,
 Macul, Santiago, Chile  \\
$^{10}$Vatican Observatory, 00120 Vatican City State, Italy \\ 
$^{11}$Departamento de Ciencias Fisicas,  Universidad Andres Bello, Republica 220, Santiago, Chile \\
$^{12}$European Southern Observatory, Karl-Schwarzschild-Strasse 2, 85748 Garching, Germany \\
$^{13}$Excellence Cluster Universe, Boltzmannstr. 2, 85748, Garching, Germany \\
$^{14}$Departamento de Fisica-Universidade Federal de Sergipe, \\Rod. Marechal Rondon s/n-Jardim Rosa Elze, Sao Cristovao, 49.100-000, Sergipe, Brazil \\ 
$^{15}$Instituto de Astrof\'isica, Facultad de F\'isica, Pontificia Universidad Cat\'olica de Chile,  Av. Vicu\~na Mackenna, Santiago, Chile  \\
$^{16}$Warsaw University Observatory, Al. Ujazdowskie 4, 00-478 Warszawa,Poland \\ 
$^{17}$Department of Astronomy, Ohio State University, 140 W. 18th Ave., Columbus, OH 43210 \\ 
$^{18}$Universidad de Concepci{\'o}n, Departamento de Astronomia,
Casilla 160--C, Concepci{\'o}n, Chile \\ 
$^{19}$ Department of Physics, University of Warwick, Gibbet Hill Road, Coventry, CV4 7AL, UK \vspace{-0.5cm}}
\begin{document}
\include{journaldefs}
\date{Accepted ...... Received ...... ; in original form......   }
\pagerange{\pageref{firstpage}--\pageref{lastpage}} \pubyear{2015}
\maketitle
\label{firstpage}
\vspace{-0.50cm}
\begin{abstract}
We investigate interstellar extinction curve variations toward $\sim$4 deg$^{2}$ of the inner Milky Way in  $VIJK_{s}$ photometry from the OGLE-III and $VVV$ surveys, with supporting evidence from diffuse interstellar bands and $F435W,F625W$ photometry. We obtain independent measurements toward $\sim$2,000 sightlines of $A_{I}$, $E(V-I)$, $E(I-J)$, and $E(J-K_{s})$, with median precision and accuracy of 2\%. We find that the variations in the extinction ratios $A_{I}/E(V-I)$, $E(I-J)/E(V-I)$ and $E(J-K_{s})/E(V-I)$ are large (exceeding 20\%), significant, and positively correlated, as expected. However, both the mean values and the trends in these extinction ratios are drastically shifted from the predictions of Cardelli and Fitzpatrick, \textit{regardless of how $R_{V}$ is varied}. Furthermore, we demonstrate that variations in the shape of the extinction curve has at least two degrees of freedom, and not one (e.g. $R_{V}$), which we conform with a principal component analysis. We derive a median value of $<A_{V}/A_{Ks}>=13.44$, which is $\sim$60\% higher than the ``standard" value. We show that the \textit{Wesenheit} magnitude $W_{I}=I-1.61(I-J)$ is relatively impervious to extinction curve variations. 

Given that these extinction curves are linchpins of observational cosmology, and that it is generally assumed that $R_{V}$ variations correctly capture variations in the extinction curve, we argue that systematic errors in the distance ladder from studies of type Ia supernovae and Cepheids may have been underestimated. Moreover, the reddening maps from the \textit{Planck} experiment are shown to systematically overestimate dust extinction by $\sim$100\%, and lack sensitivity to extinction curve variations. \vspace{-0.0cm} 
\end{abstract}
\maketitle
\begin{keywords} \vspace{0.1cm} ISM: dust, extinction  -- ISM: lines and bands \vspace{-2.0cm} \end{keywords}

\clearpage

\section{Introduction}

\subsection{Interstellar extinction curve towards the inner Milky Way}
\label{sec:Introduction}
The interstellar extinction curve toward the inner Milky Way has long been argued to be ``non-standard". This was first suggested from observations of planetary nebulae. \citet{1992A&A...266..486S} analyzed data on the Balmer decrement and the ratio of radio to H$\beta$, and argued for an extinction curve that declines more rapidly with increasing wavelength (i.e. steeper) than the $R_{V}=A_{V}/E(B-V)=3.1$ extinction curve, either $R_{V}=2.0$ or $R_{V}=2.7$ depending on the choice of parameterization, a conclusion supported by \citet{1992A&AS...95..337T} using a similar method. \citet{1994A&A...289..261P} confirm a systematic difference in extinction derived from the radio/H$\beta$ flux ratio and the Balmer decrement using additional measurements. \citet{2004MNRAS.353..796R} obtain additional measurements for a set of lines ([OIII], H$\alpha$, etc) and estimate a mean extinction curve toward the inner Galaxy of $<R_{V}>=2.0$.

Separately, and independently, the anomalous dereddened colours of standard crayons in and near the Galactic bulge such as RR Lyrae stars have also suggested a non-standard extinction curve \citep{1999ApJ...521..206S}. The evidence has since accumulated from measurements of magnitude-colour slope of red clump (RC) centroids in optical photometry (\citealt{2003ApJ...590..284U,2013ApJ...769...88N}, following the work of \citealt{1996ApJ...460L..37S} and \citealt{1996ApJ...464..233W}) from the \textit{Optical Gravitational Lensing Experiment} (OGLE), in optical \textit{Hubble Space Telescope} (HST) photometry \citep{2010A&A...515A..49R}, in ground-based near-infrared (IR) photometry \citep{2009ApJ...696.1407N}, with a combination of ground-based near-IR and space-based mid-IR photometry from the \textit{Spitzer Space Telescope} \citep{2009ApJ...707..510Z},  in optical photometry of RR Lyrae stars \citep{2012ApJ...750..169P,2015ApJ...811..113P}, with photometry of individual red giant stars in near-IR photometry \citep{2009MNRAS.394.2247G}, and from line-emission ratios toward the Galactic centre \citep{2011ApJ...737...73F}. For dissenting analyses, see \citet{2008AJ....135..631K} and \citet{2013A&A...550A..35P}. 

The mean shape and variation of the extinction curve toward the bulge is of concern to the fields of microlensing  (e.g. \citealt{2012A&A...547A..55B,2014ApJ...794...52H,2015ApJ...810..155Y}), extremely-metal-poor stars \citep{2014MNRAS.445.4241H,2015arXiv151103930H}, Galactic globular clusters \citep{2012ApJ...755L..32M,2015ApJ...806..152S}, and Galactic structure \citep{2013MNRAS.434..595C,2015MNRAS.450.4050W}, among others. \citet{2014A&A...571A..91M} argued that a large fraction of the interstellar dust toward the bulge is located in a single ``great dark lane". The level of interest and controversy has thus been amply justified and driven by the research needs. 

\citet{2013ApJ...769...88N} followed up on the issue of extinction curve variations by combining measurements of the optical reddening $E(V-I)$ from  $\sim$100 deg$^{2}$ of photometry from OGLE-III with the corresponding measurements of the near-IR reddening $E(J-K_{s})$ from the \textit{VISTA Variables in the Via Lactea} $(VVV)$ survey \citep{2012A&A...543A..13G}. The mean extinction ratios measured were $ A_{I}/E(V-I) \approx 1.22$, and $E(J-K_{s})/E(V-I) \approx 0.34$, both with statistically significant 1-$\sigma$ scatter of $\sim$10\%, and both approximately consistent in the mean with the $R_{V} = 2.5$ interstellar extinction curve from \citet{1989ApJ...345..245C}, but not consistent with the mean value of $A_{I}/E(V-I)=1.44$ measured toward the Large Magellanic Cloud (LMC) \citep{2003ApJ...590..284U}.  One of the results from \citet{2013ApJ...769...88N}, that the optical extinction could be parameterized as $A_{I} = 0.7465E(V-I)+1.3700E(J-K_{s})$, has since been confirmed by \citet{2015ApJ...811..113P} and also used by \citet{2015ApJ...808L..12K}, in their studies of bulge RR Lyrae stars. 

A legitimate concern that one can level against the findings of \citet{2013ApJ...769...88N} is that the argument is dependent on assuming that the ``standard" extinction curve is that of  \citet{1989ApJ...345..245C}. In contrast, both \citet{1999PASP..111...63F} and \citet{2007ApJ...663..320F} find substantially steeper mean Galactic extinction curves for $\lambda \gtrsim 6000\,${\AA} even if one fixes $R_{V} \approx 3.1$. Indeed, the conclusion of \citet{2008AJ....135..631K} that the interstellar extinction toward the bulge is standard, uses the curve of \citet{1999PASP..111...63F} to anchor what ``standard" means -- if \citet{2008AJ....135..631K} had relied upon the analysis of \citet{1989ApJ...345..245C} their conclusion would have been reversed. \citet{2011ApJ...737..103S} have recently demonstrated that $R_{V}=3.1$ curve from \citet{1999PASP..111...63F}  yields a much better fit to optical photometry of main-sequence turnoff stars in the northern Galactic halo than that of \citet{1989ApJ...345..245C}, a conclusion that \citet{2014MNRAS.445.4252W} confirmed using quasi-stellar objects (QSOs) as standard crayons. \citet{2014A&A...563A..15B} recently derived arguably self-consistent distances to Galactic bulge stars by assuming the extinction curve of \citet{2007ApJ...663..320F}. 

This continuing discrepancy is thus inevitable due to the large number of degrees of freedom. When one can vary not just the parameter $R_{V}$ but also the choice of parameterization (\citealt{1989ApJ...345..245C}, \citealt{1999PASP..111...63F}, etc), one has a lot of flexibility with which to fit extinction data. A compelling option with which to resolve this discrepancy is to acquire more photometry and thus more colours, and that is what we do in this investigation. Similarly to \citet{2013ApJ...769...88N}, our combination of OGLE-III and VVV photometry allows us to measure $ A_{I}/E(V-I)$ and  $E(J-K_{s})/E(V-I)$. What we also do, by matching sources between the two catalogues, is measure the reddening ratio $ E(I-J)/E(V-I)$, for which the different parameterizations and different values of $R_{V}$ yield specific and distinct predictions. 

The structure of this paper is as follows. In Section \ref{sec:Cosmology} we discuss the relevance of the inner Milky Way extinction curve to cosmology. In Section \ref{sec:Data} we describe the raw data used in this investigation. We explain our methodology in Section \ref{sec:Methodology}, the expectations from theory are stated in Section \ref{sec:TheoryExpectations}, present our results in Section \ref{sec:Results}, we compare our results to select other investigations in Section \ref{sec:Comparison}, and our conclusions in Section \ref{sec:Summary}.

\subsection{Extinction towards Type Ia Supernovae and Extragalactic Cepheids}
\label{sec:Cosmology}
\citet{2013ApJ...769...88N} argued that there may be a link between the extinction curve variations observed toward the inner Milky Way and
observations of SN Ia, toward which low values of $R_{V}$ are common (e.g. \citealt{2011A&A...529L...4C,2014ApJ...784L..12G}). The Carnegie Supernovae Project \citep{2014ApJ...789...32B} has measured densely-sampled light curves in 8-10 bandpasses covering the optical and near-IR wavelength regime, and confirm steeper-than-standard extinction curves as common -- they report a mean $R_{V}=2.15 \pm 0.16$ toward their sample.  Similarly, \citet{2015ApJ...802...20R} find that the distance dispersion toward type Ia SNe is minimised if they assume $R_{V}=1.7$ as the mean of their sample, though \citet{2015ApJ...812...31J} have demonstrated that this result may emerge from the assumption that star-forming and quiescent galaxies have identical interstellar extinction curves. 

\begin{figure}
\begin{center}
\includegraphics[totalheight=0.36\textheight]{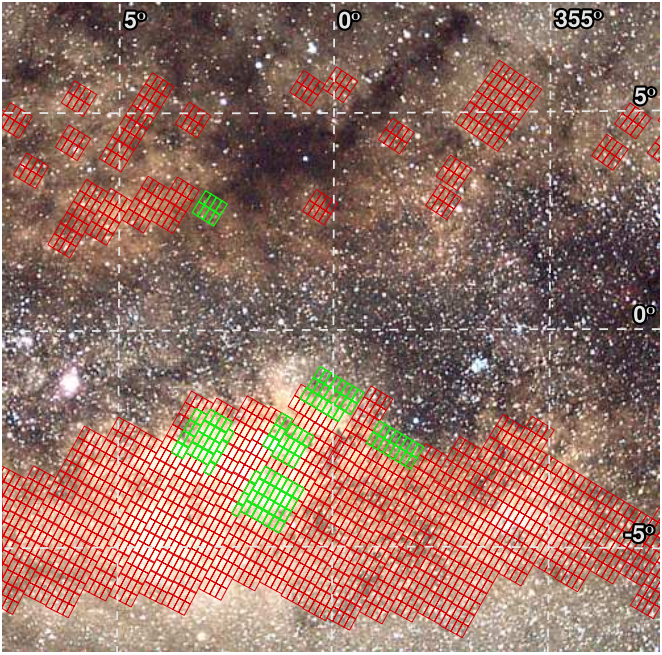}
\end{center}
\caption{\large Subset of OGLE-III subfields shown in red overlaid on an optical image of the Galactic bulge with Galactic coordinate system shown as well. The subfields used in this work, for which we also use the matching $VVV$ photometry, are shown in green.} 
\label{Fig:ogle3-blg-fields}
\end{figure}

\citet{2013ApJ...779...38P} has shown that this extinction is likely interstellar rather than circumstellar extinction, as the ratio of the equivalent width of the diffuse interstellar band at 5780 {\AA} to the inferred extinction in $V$-band is consistent with the value for the Milky Way interstellar medium (ISM). \citet{2014ApJ...789...32B} found that many of the type Ia SNe with low values of $R_{V}$ are also high-velocity events. In turn, \citet{2013Sci...340..170W} argue that high-velocity type Ia SNe are substantially more concentrated in the inner and brighter regions of host galaxies. Thus, the findings of a steeper extinction curve toward the inner Milky Way (also recently demonstrated for the Andromeda galaxy, see \citealt{2014ApJ...785..136D}) may be consistent what is found from type Ia SNe host galaxies.

These findings continue to be controversial, as the $R_{V}$ values seemingly ubiquitous toward type Ia SNe are believed to be rare in the Milky Way, and some have argued that they emerge due to a degeneracy between extinction curve variations and intrinsic colour variations of type Ia SNe \citep{2014ApJ...780...37S}. As manifestly plausible as the argument of colour variations clearly is, we suggest that some of the discrepancy is due to the following two reasons:
\begin{enumerate}
\item There is a greater range of extinction curve variations within the Milky Way than commonly believed. In particular, $R_{V}=3.1$ may simply be the most common value for the solar neighbourhood, rather than the sharply-peaked mode for the diffuse interstellar medium throughout the Galaxy.
\item The widely-used parametric fits of \citet{1989ApJ...345..245C} and  \citet{1999PASP..111...63F}, may simply be incorrect descriptions of nature at the 10\% level or greater. In an era of $\lesssim 3\%$ cosmology, this matters, and the systematic error is manifesting as spurious values of $R_{V}$. 
\end{enumerate}
The anonymous referee notes that the investigation of \citet{2007ApJ...663..320F} already addressed some of these issues. We quote directly from the referee:
\begin{quotation}
\vspace{-0.28cm} \noindent ``\citet{2007ApJ...663..320F} have demonstrated that interstellar extinction is, in general, far more complex than implied by the CCM [re: \citealt{1989ApJ...345..245C}] relations and that the apparent CCM relations are largely the result of correlated errors. Even CCM never intended the relationships to be considered a ``law", but rather a means to account for a large component of the observed variation. Further, \citet{2009ApJ...699.1209F}  have shown that optical-NIR extinction curves in the local ISM require at least 2 parameters to explain the observed variations." - The anonymous referee.
 \end{quotation}

\vspace{-0.05cm}  Uncertainties in the mean value of the interstellar extinction curve are also emerging as an issue in the determination of the Cepheid distance scale, which anchors Hubble's constant $H_{0}$. \citet{2004MNRAS.349.1344A} estimate $R_{V}=2.5$ as the best-fit extinction curve parameter toward an archival sample of extragalactic Cepheid light curves. \citet{2015MNRAS.450.3597F} used $BVRI$ photometry of Cepheids in the maser-host galaxy NGC 4258, and found that $R_{V}=4.9$ provided the best-fit. \citet{2015MNRAS.449.1171N} recently showed that the extinction toward the Cepheids in M101 (the Pinwheel galaxy) was better characterised by $ A_{I}/E(V-I) \approx 1.15$ rather than the canonical value of $ A_{I}/E(V-I) \approx 1.47$ \citep{1989ApJ...345..245C}. The situation is such that \citet{2011ApJ...730..119R,2012ApJ...745..156R} have shifted to using $H$-band observations of Cepheids to infer distances, for which the extinction is believed to be a smaller uncertainty. However, this comes at the cost of a broader PSF (since HST photometry is nearly diffraction-limited) and thus greater blending in $H$-band, in addition to greater relative flux contributions from colder red giant branch and asymptotic giant branch stars. 

In light of these factors we consider plausible the idea that better characterisation of the extinction curve toward the inner Milky Way -- an independently interesting and tractable scientific endeavour -- may facilitate superior understanding of the extinction toward both Cepheids and type Ia SNe. Though the inner Milky Way may be dismissed as less than 2\% of the sky in surface area, it represents no less than $\sim$25\% of the stellar mass of the Milky Way \citep{1995ApJ...445..716D,2013ApJ...769...88N,2015MNRAS.448..713P}, a number which accounts for the bulge alone and does not include the inner disk in which this dust is likely located. Indeed, from \citet{2013ApJ...779..115B}, we can estimate that the Milky Way's disk contains three times as much stellar mass in the Galactocentric range $4 \leq R_{GC} \leq 8$ as it does in the outer disk $8 \leq R_{GC} \leq 20$. Thus, the dust extinction we study in this work may be that which characterizes how the Milky Way would appear to outside observers, much more so than the dust extinction curve of the solar neighbourhood, and is thus pertinent to better interpreting extragalactic stellar populations. We note of a recent pre-print posted on astro-ph, which measured a mean extinction curve parameter $R_{V}=2.4$ toward a sample of 16 intermediate-redshift quasars. 

We are aware of uncertainties in the 3D distribution of dust. \citet{1980A&AS...42..251N} report $2.6 \leq A_{V} \leq 3.3$ toward the bulk of our sightlines within 1 Kpc of the Sun, which would contribute most of the extinction measured in our work and thus contradict the statement above concerning the Galactic distribution of dust. On the other hand, \citet{2014AJ....148...24S} find an extinction excess toward the Galactic center located $\sim$6 Kpc from the Sun, which \citet{2014A&A...571A..91M} interprets as a ``Great Dark Lane" for the Milky Way. Regardless of the details of how the dust is distributed, we expect the integrated sum of extinction to span a range of Galactocentric radii. 

\section{Data}
\label{sec:Data}
OGLE-III, the third phase of the Optical Gravitational Lensing Experiment, produced photometric maps of the Galactic bulge, parts of the Galactic disk, and the Magellanic Clouds. Observations were taken with the 1.3 meter Warsaw Telescope, located at the Las Campanas Observatory. The camera had eight 2048${\times}$4096 detectors, with a scale of approximately 0.26$\arcsec$/pixel yielding a combined field of view $0.6^{\circ}\times0.6^{\circ}$. More detailed descriptions of the instrumentation, photometric reductions and astrometric calibrations are available in  \citet{2003AcA....53..291U}, \citet{2008AcA....58...69U} and \citet{2011AcA....61...83S}. The photometry is in the $VI$ optical filters as calibrated by Landolt standard stars \citep{1992AJ....104..340L}. OGLE-III photometry and reddening maps are available for download from the OGLE webpage\footnote{http://ogle.astrouw.edu.pl/}. The fourth phase of the OGLE project (OGLE-IV, \citealt{2015AcA....65....1U}) has actually been underway since 2010, with photometric coverage of about 2,100 square degrees of the Galactic bulge and disk. However, we use OGLE-III photometry in this study for consistency with prior investigations. 


The \textit{VISTA Variables in the Via Lactea} (VVV) ESO public survey \citep{2010NewA...15..433M} is a near-IR photometric survey covering 560 deg$^{2}$ of the Galactic bulge and southern disk. Observations were carried with the VISTA InfraRed Camera (VIRCAM, \citealt{2006SPIE.6269E..0XD,2010SPIE.7733E..06E}), mounted at the 4.1 m telescope VISTA (Visible and Infrared Survey Telescope for Astronomy) telescope \citep{2015A&A...575A..25S}, located in its own peak at the ESO Cerro Paranal Observatory in Chile. VIRCAM has a mosaic of 16 Raytheon VIRGO 2048${\times}$2048-pixel detectors, with a scale of 0.339"/pixel. VVV observations use the stack of two slightly dithered images to produce a stacked image known as a ``Pawprint". A sequence of 6 offset ``Pawprints" is used to cover the gaps between the detectors to produce a full, nearly uniform sky coverage of 1.50 deg$^{2}$ known as a ``tile". Images are reduced, astrometrized, and stacked at the Cambridge Astronomy Survey Unit (CASU) using the VISTA Data Flow System pipeline \citep{2004SPIE.5493..401E,2004SPIE.5493..423H,2004SPIE.5493..411I}. VVV photometric catalogues at the ``tile" level are also produced at CASU. A detailed description of these VVV catalogues can be found in \citet{2012A&A...537A.107S}. For this work we use catalogues obtained using PSF photometry based on DoPHOT \citep{1993PASP..105.1342S,2012AJ....143...70A} measurements from pawprint stacked $J$ and $K_{s}$ band images. CASU photometric catalogues are then used to calibrate the PSF photometry. Final magnitudes are therefore based on the VISTA photometric system. Details on the construction of VVV PSF catalogues can be found in \citep{2015AJ....149...99A}. Infrared $E(J-K_{s})$ reddening maps measured with $VVV$ photometry can be found on the online BEAM calculator\footnote{http://mill.astro.puc.cl/BEAM/calculator.php}.

We use observations from a subset of OGLE-III fields that were selected such that our study would span the range of reddening curve parameters $E(J-K_{s})/E(V-I)$ toward the bulge, as measured by \citet{2013ApJ...769...88N}. The photometric coverage used in this work is shown in Figure \ref{Fig:ogle3-blg-fields}. For those sightlines, we match to sources identified in $VVV$ photometry, producing a combined $VIJK_{s}$ catalogue, where we leave the near-IR photometry in the original VISTA filter system and do not transform to 2MASS \citep{2006AJ....131.1163S}.  The total photometry is then broken up into nearly non-overlapping (and thus independent) rectangles, there is a small amount of overlap when different OGLE-III fields lie slightly atop one another. The rectangles into which the OGLE-III subfields are broken up into are chosen such that the number of RC stars, $N_{RC}$ within the region fit is $100 \leq N_{RC} \leq 200$. The typical angular size of these sightlines is $3'{\times}3'$.

\section{Methodology}
\label{sec:Methodology}

\subsection{Photometric Zero Points of the Red Clump}
The mean intrinsic (dereddened) colour of the Galactic bulge RC is measured \citep{2013A&A...549A.147B} to be:
\begin{equation} 
(V-I)_{RC,0}=1.06. 
\end{equation}
That is the value adopted by our investigation as the zero-point, it is derived by equating the photometric colours of the main-sequence turnoff and subgiant branch stars studied by  \citet{2013A&A...549A.147B} with their spectroscopic temperatures, as derived from high signal-to-noise, high-resolution spectra. This derivation is consistent with several other determinations:
\begin{itemize}
\item The prediction from the BaSTI isochrones \citep{2004ApJ...612..168P,2007AJ....133..468C}  for a 12 Gyr old, [Fe/H]$=0$ stellar population  is $(V-I)_{RC,0}=1.06$  (\citealt{2014MNRAS.442.2075N}, Table 1);
\item Based on empirically-calibrated population parameters \citep{2010AJ....139..329T,2011A&A...525A...2B,2012A&A...543A.106B,2014ApJ...796...68B,2015ApJ...798L..41C} and observed photometry \citep{2007AJ....133.1658S,2003PASP..115..413S} from 47 Tuc and NGC 6791, and assuming Baade's window metallicity-distribution function (MDF) \citep{2011A&A...534A..80H}, the Galactic Bulge RC value is $(V-I)_{RC,0} \approx 1.07$. This is marginally lower than that estimated by \citet{2013ApJ...769...88N} using the same method due to the increased best-fit metallicity for NGC 6791.
\item The Galactic bulge RC is 0.55 mag redder than Galactic bulge ab-type RR Lyrae \citep{2013ApJ...769...88N}, where throughout this discussion we only refer to ab-type RR Lyrae. The Fourier coefficients of the RR lightcurves yield a mean intrinsic RR Lyrae colour of $(V-I)_{RR}=0.49$ \citep{2015ApJ...811..113P}, for an estimated mean intrinsic RC colour of $(V-I)_{RC,0}=1.04$. Alternatively, we have selected 2,301 RR Lyrae from OGLE-III with four or more $V$-band observations near minimum light (phase $0.50 \leq \phi \leq 0.78$), and contrasted them to the nearest RC reddening measured by \citep{2013ApJ...769...88N}. We kept the data points where the reddening agreed to be better than 0.20 mag in $E(V-I)$ to remove spurious outliers, and for which $(V-I)_{RC} \leq 3.30$, leaving 1,987 RR Lyrae. We find that the ratio of reddening agrees to better than 1\% in the mean, and that the difference in reddening is 0.02 mag when we regress the offset to $E(V-I)_{RC}=0$ -- exactly equivalent to the above inferred value of $(V-I)_{RC,0}=1.04$;
\item The near-IR colour $(J-K_{s})_{RC,0}=0.68$ was derived by \citet{2011A&A...534A...3G},  based on a determination of $E(B-V)=0.55$ toward Baade's window by \citet{2008A&A...486..177Z}.  Applying the conversions from Table \ref{table:1}, which is explained below, this corresponds to $(V-I)_{RC,0} \approx1.07$;
\item The reddening measurements of \citet{2011A&A...534A...3G} were in turn tested by comparing spectroscopic temperatures and photometric temperatures \citep{2014A&A...569A.103R}. The discrepancy in the zero-point is measured to be no greater than ${\Delta}E(J-K_{s}) = -0.006  \pm 0.026$ -- consistent with zero;
\end{itemize}

Our assumed mean colour $(V-I)_{RC,0}=1.06 \pm 0.03$ is thus well-supported by a diverse array of inferential methods, where the 0.03 mag is a conservative estimate of the 1-$\sigma$ error from the arguments presented above. A possible concern is that the effect of metallicity variations can be significant: interpolating between the BaSTI-predicted values for [Fe/H]$=-0.35$ and [Fe/H]$=+0.40$ yields a derivative of $d(V-I)_{RC}/d\rm{[Fe/H] }\approx 0.29$ mag dex$^{-1}$  (\citealt{2014MNRAS.442.2075N}, Table 1), with the effect of variations in age or [$\alpha$/Fe] being negligible.  However, the total range in the mean metallicity across our fields are no greater than $\sim$0.10 dex (\citealt{2013A&A...552A.110G}, see also \citealt{2012ApJ...746...59R})  and thus the effect is negligible. This emerges due to our choice of sightlines, which are relatively similar in direction, do not span the whole bulge, and thus do not probe a significant spread in mean metallicity. 

\begin{table}
\centering
\begin{tabular}{|l|rr|}
	\hline \hline
Index & Colour 1 & Colour 2  \\
	\hline \hline \hline
$(V-I)$ & 1.060 & 0.994 \\
$(V-U)$ & $-$2.217 & $-$1.746 \\
$(V-B)$ & $-$1.115  & $-$0.982 \\
$(V-R)$ & 0.568  & 0.523 \\
$(V-Z_{\rm{Vista}})$ & 1.313 & 1.220  \\
$(V-Y_{\rm{Vista}})$ & 1.531 & 1.428 \\
$(V-J_{\rm{Vista}})$ & 1.893 & 1.773 \\
$(V-H_{\rm{Vista}})$ & 2.361 & 2.220 \\
$(V-K_{s,\rm{Vista}})$ & 2.518 & 2.355 \\  
$(V-J_{\rm{2MASS}})$ & 1.852 & 1.734 \\
$(V-H_{\rm{2MASS}})$ & 2.378  & 2.236 \\
$(V-K_{s,\rm{2MASS}})$ & 2.525 & 2.361 \\  
$(V-u_{\rm{SDSS}})$ & $-$3.104 & $-$2.620  \\  
$(V-g_{\rm{SDSS}})$ & $-$0.553 &  $-$0.476 \\  
$(V-r_{\rm{SDSS}})$ & 0.327 &  0.288 \\  
$(V-i_{\rm{SDSS}})$ & 0.600 &  0.543 \\  
$(V-z_{\rm{SDSS}})$ & 0.759 &  0.683 \\  
$(V-F435W_{\rm{ACS}})$ & $-$1.161 & $-$1.017  \\  
$(V-F475M_{\rm{ACS}})$ & $-$0.623 & 0.554 \\  
$(V-F555W_{\rm{ACS}})$ & $-$0.046 & $-$-0.040 \\  
$(V-F606W_{\rm{ACS}})$ & 0.286 & 0.262 \\  
$(V-F814W_{\rm{ACS}})$ & 1.079 & 1.015 \\  
$(V-F218W_{\rm{WFC3}})$ & $-$6.929 & $-$6.576 \\  
$(V-F225M_{\rm{WFC3}})$ & $-$6.820 & $-$6.118 \\  
$(V-F275W_{\rm{WFC3}})$ & $-$5.192 & $-$4.300 \\  
$(V-F336W_{\rm{WFC3}})$ & $-$2.387 & $-$1.845 \\  
$(V-F350lp_{\rm{WFC3}})$ & 0.1430 & 0.127 \\  
$(V-F390m_{\rm{WFC3}})$ & $-$2.490 & $-$2.087 \\  
$(V-F390W_{\rm{WFC3}})$ & $-$1.749 & $-$1.458 \\  
$(V-F438W_{\rm{WFC3}})$ & $-$1.192 & $-$1.036 \\  
$(V-F475W_{\rm{WFC3}})$ & $-$0.588 & $-$0.522  \\  
$(V-F547M_{\rm{WFC3}})$ & 0.004 & 0.007 \\  
$(V-F555W_{\rm{WFC3}})$ & $-$0.103 & $-$0.092 \\  
$(V-F606W_{\rm{WFC3}})$ & 0.271 & 0.248 \\  
$(V-F625W_{\rm{WFC3}})$ & 0.491 & 0.451 \\  
$(V-F775W_{\rm{WFC3}})$ & 0.993 & 0.935 \\  
$(V-F814W_{\rm{WFC3}})$ & 1.070 & 1.006 \\  
$(V-F850LP_{\rm{WFC3}})$ & 1.284 & 1.205 \\  
$(V-F098M_{\rm{WFC3}})$ & 1.380 & 1.292 \\  
$(V-F110W_{\rm{WFC3}})$ & 1.657 & 1.549 \\  
$(V-F125W_{\rm{WFC3}})$ & 1.845 & 1.727 \\  
$(V-F140W_{\rm{WFC3}})$ & 2.070 & 1.946 \\  
$(V-F160W_{\rm{WFC3}})$ & 2.279 & 2.152 \\  
\hline
$(I-J_{\rm{Vista}})$ & 0.833 & 0.779 \\
$(J_{\rm{Vista}}-K_{s, \rm{Vista}})$ & 0.653 & 0.582  \\
	\hline
\end{tabular}
\caption{Estimated mean photometric colours for the Galactic bulge red clump. In the second column (Colour 1), we assume a model atmosphere with $\log{g}=2.2$, [Fe/H]$=0.0$,  [$\alpha$/Fe]$=0.0$, and $T_{\rm{eff}}=4650$K, which yield the intrinsic red clump colours adopted in this work. In the third column (Colour 2), we assume $\log{g}=2.2$, [Fe/H]$=-0.30$,  [$\alpha$/Fe]$=+0.10$, and $T_{\rm{eff}}=4800$K. Colours are computed using the methodology of \citet{2014MNRAS.444..392C}, where the accuracy of synthetic colors is also discussed (in particular the shortcomings at blue and ultraviolet wavelengths). $UBVRI$ and $WFC3$ magnitudes are in the Vega system, $SDSS$ magnitudes are in the $AB$ system. Some of the colours are thus in a composite Vega-AB system.    }
\label{table:1}
\end{table}
\vspace{-0.00in}


We use model atmospheres to estimate the intrinsic colours of the bulge RC in the full range of filters used in this work, as well as others that may be of interest to future studies, which we show in Table \ref{table:1}. The atmospheric parameters $\log{g}=2.2$ and [Fe/H]$=0$ are typical of the RC \citep{2013MNRAS.430..836N}, and $T_{\rm{eff}}=4650$K is chosen to agree with the intrinsic colour $(V-I)_{RC,0}=1.06$. The remaining synthetic colours were computed interpolating over a grid of MARCS model atmosphere \citep{2008A&A...486..951G} at the $T_{\rm{eff}}$, $\log{g}$ and [Fe/H] quoted above, and appropriate filter transmission curves (see details in \citealt{2014MNRAS.444..392C} ). Synthetic optical and 2MASS $JHK$ magnitudes were transformed into the VISTA system\footnote{see \url{http://casu.ast.cam.ac.uk/surveys-projects/vista/technical/photometric-properties} for more information on the VISTA photometric system.}. The remaining colours emerge from a detailed model atmosphere calculation. Optical colours were computed directly We also provide the colour determinations for a star with atmospheric parameters $\log{g}=2.2$, [Fe/H]$=-0.30$,  [$\alpha$/Fe]$=+0.10$, and $T_{\rm{eff}}=4800$K. These two model atmospheres together define the vector over which metallicity gradients (and thus temperature gradients, since the colour of the RC is predominantly a function of metallicity) would matter if a study such as this one was extended over a larger swath of the bulge.

For the dereddened apparent magnitude of the RC, we use the equation:
\begin{equation}
I_{RC,0} = 14.3955 - 0.0239*l + 0.0122*|b|,
\end{equation}
where the zero-point is taken from \citet{2013ApJ...769...88N}, and the derivatives are derived by fitting to the data of \citet{2013MNRAS.435.1874W} within the coordinate range $-2.00^{\circ} \leq  l \leq 4.0^{\circ},  |b| \leq 4.5^{\circ}$. The gradients emerge due to projection effects, as stars toward the Galactic bulge are distributed as a bar. The standard deviation between the fit and the data is 0.018 mag. We have verified that there is no significant evidence for cross-terms or higher-order terms in longitude or latitude for the apparent magnitude. Our zero-point assumption of the apparent magnitude of the RC is, as per the work of \citet{2013MNRAS.434..595C}, equivalent to assuming:
\begin{equation}
M_{I,RC} = -0.12 - 5\log{(R_{0}/8.13)},
\label{EQ:ClumpMagnitude}
\end{equation}
where $R_{0}=R_{GC,\odot}$ is the distance between the Sun and the Galactic centre. Equation \ref{EQ:ClumpMagnitude} is consistent with the canonical Galactocentric distance of $8.33 \pm 0.11$ Kpc \citep{2015MNRAS.447..948C} if $M_{I,RC}=-0.17\pm0.03$. From these arguments, the zero-point systematic error (bias) in our extinctions is exactly 0 if $M_{I,RC}=-0.17$ and $R_{0}=8.33$. However, \citet{2013ApJ...766...77N} estimated $M_{I,RC}=-0.12$ by means of an empirical calibration,  and \citet{1998ApJ...503L.131S} measured $M_{I,RC}=-0.23$ for the solar neighbourhood RC. We thus assume a 1-$\sigma$ systematic error of 0.05 mag in the values of $A_{I}$. 


\subsection{The Measurement Errors in the Reddening and Extinction}
\label{sec:Errors}
From the arguments in the preceding section, the zero-point (systematic) errors in $A_{I}$, $E(V-I)$, $E(I-J)$ and $E(J-K_{s})$ are no greater than 0.05 mag, 0.03 mag, 0.03 mag, and 0.02 mag respectively, where the errors on $E(I-J)$ and $E(J-K_{s})$ are derived from the error in $E(V-I)$ and the vectors defined by Table \ref{table:1}. These are errors which will shift the entire scale, with the errors in $E(V-I)$, $E(I-J)$ and $E(J-K_{s})$ being positively correlated. The total colour error  due to mean metallicity variations in our sample goes as $\sim1/\sqrt(12){\times}$ the spread due to metallicity if we use the uniform distribution as a probabilistic proxy, and thus the error from metallicity variations is 0.006 mag, 0.006 mag, 0.005 mag, and 0.007 mag respectively. The systematic errors in the colours are positively-correlated with each other, and negatively correlated with the error in the brightness, since the RC becomes redder and dimmer with increasing metallicity. 

The statistical errors in the fit, due to the finite number of RC stars per sightline, average 0.038 mag in $A_{I}$, and 0.011 mag in $E(V-I)$. The corresponding errors in $E(I-J)$ and $E(J-K_{s})$ are likely smaller than and positively correlated with the error in $E(V-I)$, since both the reddening and the dispersion due to temperature are smaller at these longer wavelengths. The correlation in the statistical error with $A_{I}$ is virtually zero, as the red giant branch is nearly vertical in the $VI$ colour-magnitude diagram (CMD) at the location of the RC. 

These errors are small relative to both the mean values and sample dispersions of $A_{I}=1.91 \pm 0.67$, $E(V-I) = 1.61 \pm 0.60$, $E(I-J)=1.18 \pm 0.40$ and $E(J-K_{s}) = 0.47 \pm 0.17$ in our sample, where the measured sample means and sample dispersions are discussed in Section \ref{sec:Results}. We thus measure highly significant extinction and reddening values, with mean accuracies no worse than 3\%, 2\%, 3\%, and 4\% respectively, and mean precisions no worse than 2\%, 1\%, 1\%, and 2\% respectively. Given that the errors in our reddening values are positively-correlated with one another, the errors in the fractions $E(I-J)/E(V-I)$ and $E(J-K_{s})/E(V-I)$ will be even smaller than the errors in the constituent parts. 

\subsection{Fitting for the Red Clump Magnitude and Colours}
\label{sec:Fitting}
The iterative fit for the RC apparent magnitude assumes the same luminosity function as used by \citet{2013ApJ...769...88N}:
\begin{equation}
\begin{split}
N(I)dI = A\exp\biggl[B(I-I_{RC})\biggl] + \\
\frac{N_{RC}}{\sqrt{2\pi}\sigma_{RC}}\exp \biggl[{-\frac{(I-I_{RC})^2}{2\sigma_{RC}^2}}\biggl] +  \\
  \frac{N_{RGBB}}{\sqrt{2\pi}\sigma_{RGBB}}\exp \biggl[{-\frac{(I-I_{RGBB})^2}{2\sigma_{RGBB}^2}}\biggl]   + \\\frac{N_{AGBB}}{\sqrt{2\pi}\sigma_{AGBB}}\exp \biggl[{-\frac{(I-I_{AGBB})^2}{2\sigma_{AGBB}^2}}\biggl].
\end{split}
\end{equation}
Further discussion and details on the issues pertaining to this fit, such as how to account for the red giant branch bump (RGBB) and asymptotic giant branch bump (AGBB), are by now well-documented in the literature \citep{2011ApJ...730..118N,2014MNRAS.442.2075N,2015MNRAS.447.1535N,2011ApJ...735...37C,2011A&A...534A...3G,2012A&A...543A..13G,2013MNRAS.435.1874W,2015MNRAS.450.4050W}. We assume the same stellar parameters for the RGBB and AGBB as \citet{2013ApJ...769...88N}.

\begin{figure}
\begin{center}
\includegraphics[totalheight=0.33\textheight]{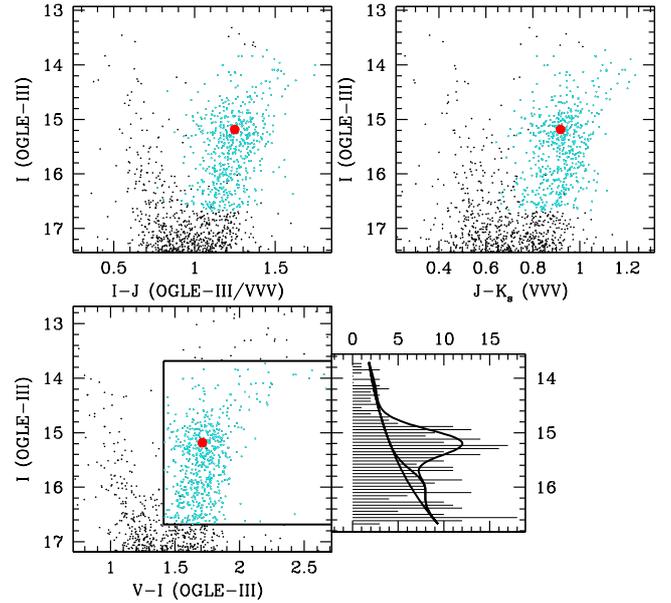}
\end{center}
\caption{\large Colour-magnitude diagrams in $I$ vs $(I-J,J-K_{s},V-I)$ toward $(l,b)=(2.94^{\circ},+3.06^{\circ})$, with a luminosity function and best-fit model in the histogram. The cyan dots are those over which the fit is performed, and the black dots are the entire detected colour-magnitude diagram. The three measured colours of the red clump, delineated by the red dots, are well-measured in all three colour-magnitude diagrams. } 
\label{Fig:MarkovPresentationPlot2}
\end{figure}

\begin{figure}
\begin{center}
\includegraphics[totalheight=0.33\textheight]{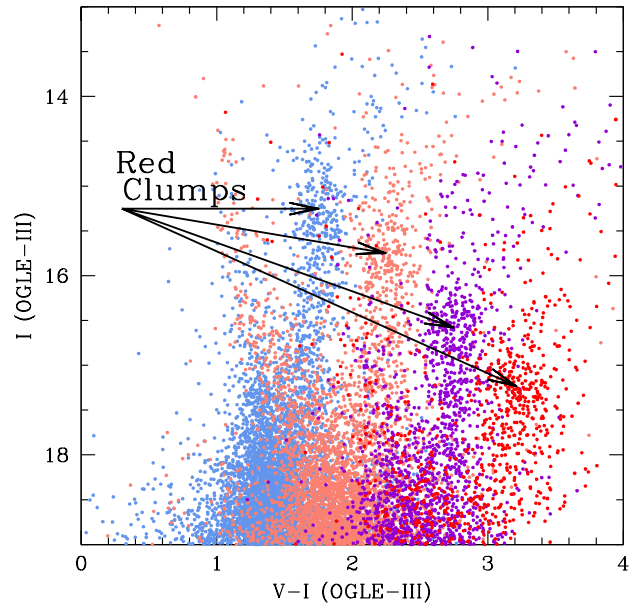}
\end{center}
\caption{\large $I$ vs $(V-I)$ colour-magnitude diagrams of four red giant branches with red clump colours of $(V-I)_{RC}=$1.76, 2.25, 2.75, and 3.22. Reddening and extinction are clearly quantities that can be precisely measured. } 
\label{Fig:MosaicCMD}
\end{figure}

We require a dual colour-cut and a magnitude cut on our sample to select the red giant branch:
\begin{equation}
\begin{split}
(V-I) \geq (V-I)_{RC}-0.30, \\
(J-K_{s}) \geq (J-K_{s})_{RC}-0.30, \\
|I-I_{RC}| \leq 1.50,
\end{split}
\end{equation}
where the colour-cuts are only applied if the colours are measured. The fit is repeated until the guessed parameters agree with the output parameters\footnote{The condition is relaxed to 0.04 mag if $(V-I)_{RC} \geq 3.20$. } to 0.03 mag in $(V-I)$ and 0.10 mag in $I$. The relatively weak Gaussian priors in the parameters $A$, $B$, and $N_{RC}$:
\begin{equation}
\begin{split}
B \sim \mathcal{N}(0.55,0.03) \\
N_{RC}/A \sim \mathcal{N}(1.17,0.07),
\end{split}
\end{equation}
are imposed to increase stability of the fit and, thus, reduce the scatter in the derived value of $I_{RC}$. The three RC colours $(V-I)_{RC}$, $(I-J)_{RC}$, and $(J-K_{s})_{RC}$ are determined independently, by picking the colour that minimizes the dispersion in colour at the luminosity of the RC.  A demonstration of the colour and magnitude determinations is shown in Figure \ref{Fig:MarkovPresentationPlot2}.  We show four fields with four different reddening values in Figure \ref{Fig:MosaicCMD}.

\begin{table*}
\centering
\begin{tabular}{|l|lllllll|}
	\hline \hline
 &  C89 &  C89 &  C89  &  F99  &  F99 &  FM07  & FM07 (RRab) \\ 
  & $R_{V}=3.1$ & $R_{V}=3.1$ & $R_{V}=2.5$   & $R_{V}=3.1$  & $R_{V}=2.5$ & $R_{V}=3.001$ & $R_{V}=3.001$ \\ 
  -- & $A_{V}=3$  & $A_{V}=6$ & $A_{V}=4$ & $A_{V}=4$  & $A_{V}=4$  & $A_{V}=4$  & $A_{V}=4$ \\
	\hline \hline \hline
$A_{V}/A_{5500}$ & 0.986 & 0.978 & 0.980 & 0.976 & 0.967 & 0.970 & 0.985  \\
$A_{I}/A_{5500}$ & 0.588 & 0.583 & 0.542 & 0.557 & 0.524 & 0.524 & 0.527 \\
$A_{J}/A_{5500}$ & 0.282 & 0.281 & 0.253 & 0.262 & 0.265 & 0.231 & 0.232 \\
$A_{K_{s}}/A_{5500}$ & 0.119 & 0.119 & 0.107 & 0.117 & 0.120 & 0.087 & 0.087 \\
	\hline
\end{tabular}
\caption{Extinction coefficients as a function of parameterization, $R_{V}$, $A_{V}$, and input stellar spectra for a few representative cases. The RR Lyrae spectrum assumes $T_{\rm{eff}}=6000$ K, [Fe/H]$=-1.0$, [$\alpha$/Fe]$=+0.40$, and $\log{g}=2.0$.  }
\label{table:2}
\end{table*}
\vspace{-0.00in}

A difficulty with our study, not shared by most previous photometric bulge studies, are the different sensitivities of the $VVV$ and $OGLE-III$ datasets. The near-IR dataset probes further down the luminosity function in highly-reddened fields. \citet{2013ApJ...769...88N} excluded sightlines with $(V-I)_{RC} \geq 3.30$ for that reason, as V magnitudes of stars located in the color-magnitude diagrams close to the RC were at or below the detection limit in OGLE dataset for reddening values $(V-I)_RC\geq 3.30$. In this investigation, in order to be able to include also such highly reddened sightlines, we fit the $(V-I)$ vs $(I-K_{s})$ colour-colour relations for stars slightly brighter and redder than the RC, satisfying $0.50<=I_Rc-I<=2.0$ and $0 \leq (I-K_{s})-(I-K_{s})_{RC} \leq 0.70$. The fit is only applied to stars redder than the RC (observationally, not intrinsically), to avoid contamination from foreground disk stars that would have lower mean reddening, and thus shifted colour-colour terms. The intercept to the colour-colour relations is used for sightlines where the intercept satisfied $(V-I)_{RC} \geq 3.20$. There is a systematic shift of 0.0413 mag between the intercept to the colour-colour relations and the RC colour determined in the standard way, plausibly due to a gravity term in the colour-colour relations. This shift is measured from sightlines where $2.40 \leq (V-I)_{RC} \leq 3.20$, and applied to more reddened sightlines,  $(V-I)_{RC} \geq 3.20$.

\section{The extinction curve in $VIJK_{s}$: Theoretical expectations}
\label{sec:TheoryExpectations}
In this section we list the predicted extinction coefficients of \citet{1989ApJ...345..245C}, \citet{1999PASP..111...63F}, and \citet{2007ApJ...663..320F}, given the filter transmission function of the photometric systems studied in this work, a RC model atmosphere, and typical extinction values. For the extinction curve of \citet{2007ApJ...663..320F}, we only show the $R_{V}=3.001$ which is their mean Galactic extinction curve -- no general formalism is provided in that work for capturing the effect of $R_{V}$ variations across the full wavelength range, deliberate on the part of the authors. 

In Table \ref{table:2} we show the predicted extinction coefficients as per a variety of assumptions. We list the broadband extinctions as ratios relative to $A_{5500}$, which is the hypothetical extinction one would measure in a narrow-band filter placed at 5500 \AA (very close but not identical to the Landolt $V$-band filter), a definition chosen to avoid ambiguities with respect to $A_{V}$ or $E(B-V)$. We find that what affects the extinction coefficients a great deal are the underlying parameterization (i.e. the chosen reference) and the choice of $R_{V}$ value. The convolution with the extinction curve itself has little impact, and thus extinction coefficients can be assumed to be independent of extinction. The ratio $E(V-I)/A_{5500 \AA}$, where $A_{5500}$ is the extinction at 5500 \AA, is predicted to shrink by $\sim$0.75\% or 0.003 as $A_{V}$ is doubled. Though there are contexts where this will matter, an offset of $\sim$0.01 mag in $E(V-I)$ is to small to affect any of the conclusions reached in this work.  We also list the predicted extinction coefficients for the spectra of a typical RR Lyrae star, they are nearly identical to those of RC stars, and thus studies of RR Lyae stars and RC stars should yield consistent answers for the photometric filters used here.

\section{Results}
\label{sec:Results}
\begin{figure}
\begin{center}
\includegraphics[totalheight=0.3\textheight]{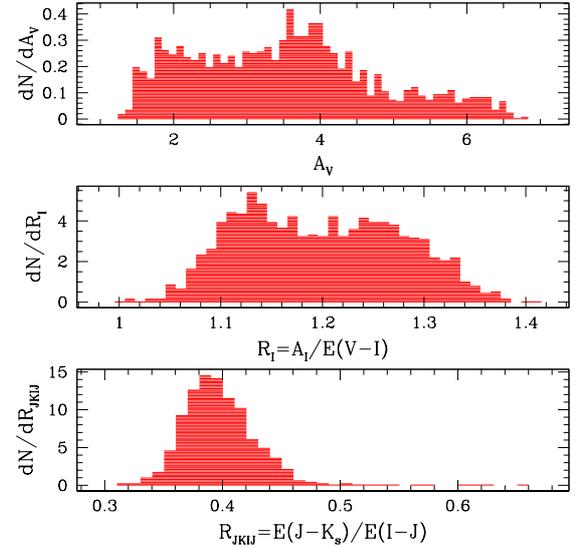}
\end{center}
\caption{\large Distributions of  the extinction $A_{V}$ and the reddening ratios $A_{I}/E(V-I)$ and $E(J-K_{s})/E(I-J)$ as measured in our study. } 
\label{Fig:ReddeningHist}
\end{figure}

We show in Figure \ref{Fig:ReddeningHist} the distribution of the extinction $A_{V}$, and the reddening ratios $A_{I}/E(V-I)$ and $E(J-K_{s})/E(I-J)$ as measured toward our 1,854 sightlines satisfying each of the photometric completeness criterion $(V-I)_{RC} \leq 4.30$, and the two differential reddening criteria $\sigma_{I,RC} \leq 0.30$ and $\sigma_{(V-I)_{RC}} \leq 0.18$. We obtain a broad distribution in each of these parameters, demonstrating our sensitivity to variations in the input parameter space. Specific findings are discussed below. The full list of values derived is available in Table \ref{table:Comprehensive}. 

\begin{table*}
\centering
\begin{tabular}{|cc|cccc|}
	\hline \hline
RA & DEC & $E(V-I)$ & $A_{I}/E(V-I)$ & $E(I-J)/E(V-I)$ & $E(J-K_{s})/E(V-I)$ \\
	\hline \hline \hline
268.257367 & -32.138720 & 1.566 & 1.243 & 0.775 & 0.308 \\ 
268.435012 & -32.119689 & 1.598 & 1.340 & 0.803 & 0.317 \\ 
268.523834 & -32.119689 & 1.430 & 1.248 & 0.799 & 0.309 \\
268.301778  &-32.119689 & 1.757 & 1.275 & 0.813 & 0.308 \\
268.212956 & -32.119689  &1.522 & 1.272 & 0.772 & 0.308  \\
	\hline
\end{tabular}
\caption{Representative sampling of the coordinates and extinction curve parameters for the 1,854 sightlines of this study deemed reliable, as per the criteria stated at the top of Section \ref{sec:Results}. Full table available as online material.}
\label{table:Comprehensive}
\end{table*}
\vspace{-0.00in}

\subsection{The Extinction Curve is Variable}
We confirm a result of \citet{2003ApJ...590..284U}, \citet{2009MNRAS.394.2247G}, and \citet{2013ApJ...769...88N}, that the extinction curve toward the inner Milky Way is variable. 

\begin{figure}
\begin{center}
\includegraphics[totalheight=0.4\textheight]{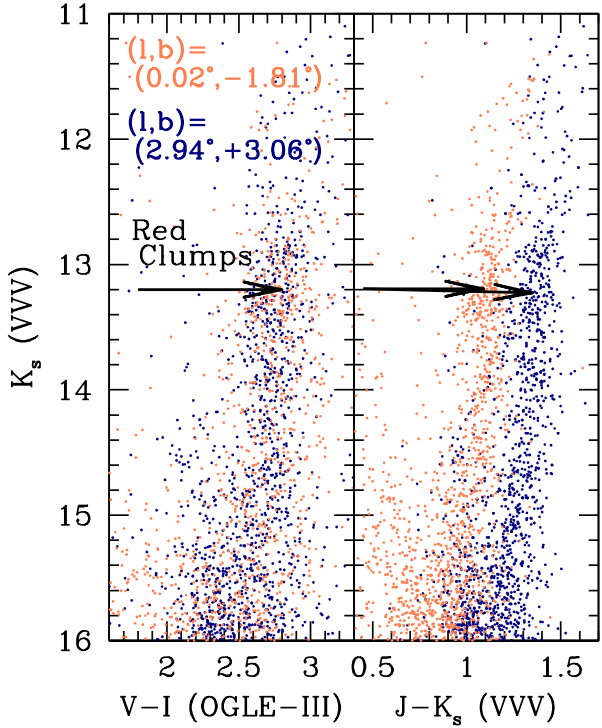}
\end{center}
\caption{\large Colour-magnitude diagrams of two bulge sightlines demonstrating the variable extinction curve. The $(V-I)_{RC}$ colours of the two sightlines agree to $\sim$0.02 mag, yet their $(J-K_{s})_{RC}$ differ by $\sim$0.28 mag, thus demonstrating a variation in the extinction curve. The extinction toward one sightline goes as $E(J-K_{s})/E(V-I) = 0.26$ (orange dots) , whereas that toward the other goes as $E(J-K_{s})/E(V-I) = 0.41$ (blue dots).   } 
\label{Fig:DoubleCMD}
\end{figure}

In the left panel Figure \ref{Fig:DoubleCMD}, we show the CMDs for two sightlines which have $E(V-I)$ values that agree to $\sim$0.02 mag, suggesting that they have the ``same reddening". They do not, that similarity in $E(V-I)$ is due to a fortuitous cancellation between the types and quantities of dust toward those two sightlines. Though the $E(V-I)$ values agree, the $E(J-K_{s})$ values differ by 0.28 mag, corresponding to 65\%. These variations are significant, and large, and thus need to be accounted for in any rigorous study of bulge photometric temperatures, metallicities, or other stellar parameters. The differences in colour cannot be due to differences in the intrinsic stellar populations, as these two sightlines have similar metallicities \citep{2013A&A...552A.110G}, whereas the right panel of Figure \ref{Fig:DoubleCMD} shows the $(J-K_{s})$ colour distribution of the red giant stars differing not just in the mean, but in fact are completely non-overlapping. 

\begin{figure}
\begin{center}
\includegraphics[totalheight=0.29\textheight]{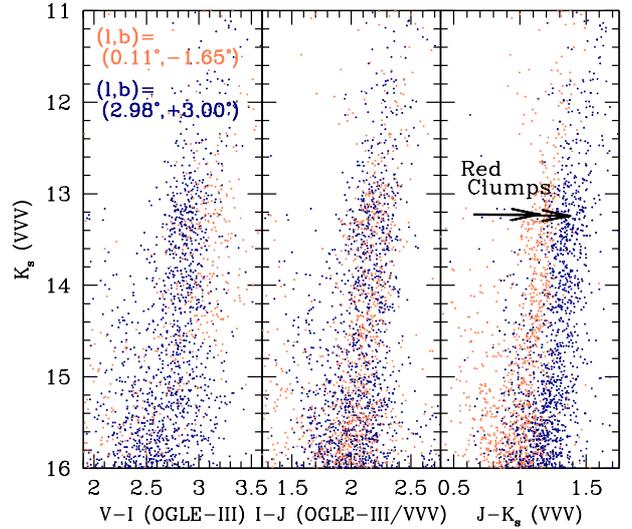}
\end{center}
\caption{\large Colour-magnitude diagrams of two bulge sightlines demonstrating the variable extinction curve. The two sightlines are selected to have reddening $E(I-J)$ that agrees to 0.01 mag. However, the sightline with lower reddening $E(V-I)$ by 0.17 mag (blue dots) has higher reddening $E(J-K_{s})$ by 0.21 mag. } 
\label{Fig:TripleCMD}
\end{figure}

We present the same idea in a different manner in Figure \ref{Fig:TripleCMD}. These two sightlines are selected to have nearly equal values of $E(I-J)$, but the sightline with greater $E(J-K_{s})$, by 0.21 mag,  has an $E(V-I)$ value that is 0.17 mag lower.


\subsection{The Extinction Curve is Non-Standard}
\begin{figure}
\begin{center}
\includegraphics[totalheight=0.36\textheight]{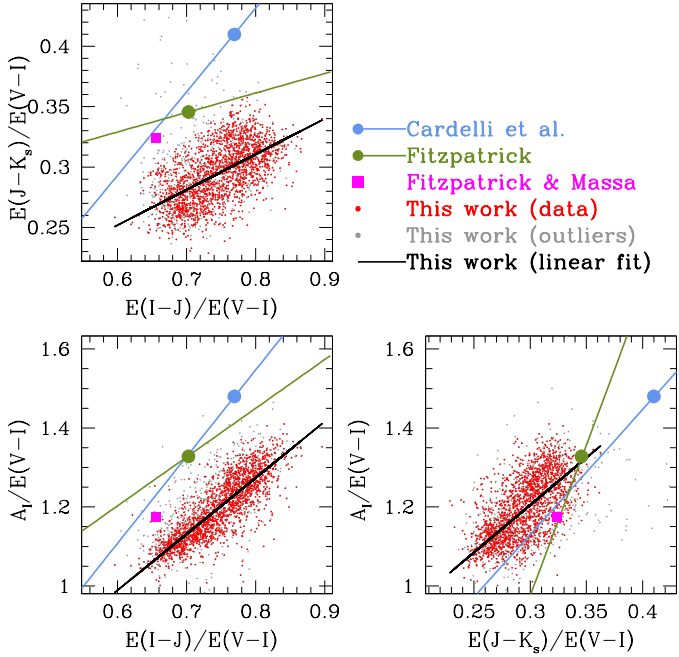}
\end{center}
\caption{\large Scatter plots of  $A_{I}/E(V-I)$, $E(I-J)/E(V-I)$, and $E(J-K_{s})/E(V-I)$ versus one another.  The extinction curves of \citet{1989ApJ...345..245C},   \citet{1999PASP..111...63F}, and \citet{2007ApJ...663..320F} are poor fits to the data both in the mean and in the trend, regardless of how $R_{V}$ is varied, with the large blue, green, and magenta dot referring to the predicted $R_{V}=3.1,3.1,3.0$ cases respectively.} 
\label{Fig:ReddeningCoefficients}
\end{figure}

We show in Figure \ref{Fig:ReddeningCoefficients} the scatter of $A_{I}/E(V-I)$, $E(I-J)/E(J-K_{s})$, and $E(J-K_{s})/E(V-I)$ relative to one another. These extinction coefficients vary in a correlated manner. We also show the predictions of  \citet{1989ApJ...345..245C},   \citet{1999PASP..111...63F}, and \citet{2007ApJ...663..320F}, which are obtained by convolving their extinction curves with a synthetic red clump atmospheric spectrum and 4 magnitudes of extinction at 5500 \AA, typical of the sightlines investigated in this work. 

The comparison to predictions leads to a conclusion that are entirely new to this investigation. Not only the $R_{V}=3.1$ curves of  \citet{1989ApJ...345..245C} and  \citet{1999PASP..111...63F} (delineated by the blue and green circles respectively) poor fits to the data over the entire span of extinction curves measured toward the bulge, but these parameterizations actually fail to intersect the bulge extinction trends \textit{regardless of how $R_{V}$ is varied.} \citet{2013ApJ...769...88N} claimed that the $R_{V}\approx 2.5$ extinction curve from \citet{1989ApJ...345..245C} was a good fit, as it nearly fit the mean values of $A_{I}/E(V-I)$ and $E(J-K_{s})/E(V-I)$, see the bottom-right panel of Figure \ref{Fig:ReddeningCoefficients}. However, the addition of the measurement $E(I-J)$ shows that the extinction curves of  \citet{1989ApJ...345..245C} (and of \citealt{1999PASP..111...63F}) fail for all values of $R_{V}$, not just in the mean, but they fail completely. The blue and green lines \textit{never} intersect the cloud of red points. 

A possible explanation for this is that  the extinction curve toward the inner Milky Way is in fact standard, but the ``standard" is not accurately characterized by the works of \citet{1989ApJ...345..245C} and \citet{1999PASP..111...63F}, and that studies of the bulge should instead use the mean Galactic extinction curve of \citet{2007ApJ...663..320F} (the magenta square in Figure \ref{Fig:ReddeningCoefficients}), which benefits from broader and more accurate measurements. The parameterization of \citet{2007ApJ...663..320F} is used by \citet{2014A&A...563A..15B} in their derivation of probabilistic distances to bulge stars. 

The mean Galactic extinction curve of  \citet{2007ApJ...663..320F} does in fact fare better. However, it is still significantly off the relations. The discrepancy corresponds to an underestimate of 0.016 (1.3\%) in the mean value of $A_{I}/E(V-I)$, an underestimate of 0.089 (12\%) in  the mean value of $E(I-J)/E(J-K_{s})$, and an overestimate of 0.026 (8.9\%) in the mean value of $E(J-K_{s})/E(V-I)$. However, the mean Galactic extinction curve of \citet{2007ApJ...663..320F}  never intersects the trend spanned by the red points, and the offsets will clearly often be larger than the offset to the mean. 

Thus, the extinction toward the inner Milky Way, both the mean curve and the dominant trends in the curve, is not well-fit by the works of   \citet{1989ApJ...345..245C},   \citet{1999PASP..111...63F}, and \citet{2007ApJ...663..320F}, even allowing for variations in $R_{V}$. 


\subsection{Whither $R_{V}$: The Shape of the Extinction Curve Has At Least Two Degrees of Freedom}

\begin{figure}
\begin{center}
\includegraphics[totalheight=0.17\textheight]{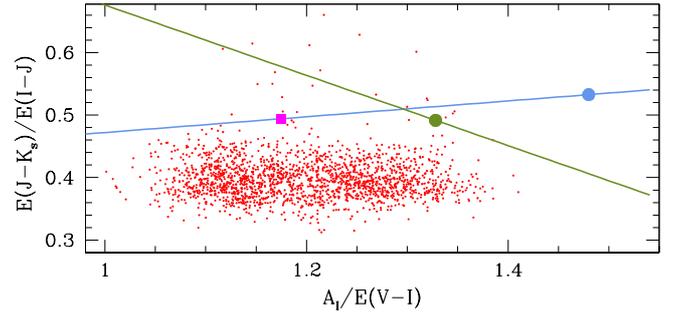}
\end{center}
\caption{\large Scatter plots of $E(J-K_{s})/E(I-J)$ versus $A_{I}/E(V-I)$.  The extinction curves of \citet{1989ApJ...345..245C},   \citet{1999PASP..111...63F}, and \citet{2007ApJ...663..320F} are poor fits to the data both in the mean and in the trend, regardless of how $R_{V}$ is varied.   } 
\label{Fig:ReddeningCoefficients2}
\end{figure}

We show in Figure \ref{Fig:ReddeningCoefficients2} the distribution of $E(J-K_{s})/E(I-J)$ vs $A_{I}/E(V-I)$ -- they appear uncorrelated. A pearson coefficient for 1,854 measurements satisfying the criteria $(V-I)_{RC} \leq 4.30$, $\sigma_{(V-I)_{RC}} \leq 0.18$, $R_{I} \leq 1.45$ and $\sigma_{I,RC} \leq 0.30$ yields $\rho= -0.0274$ -- effectively zero. As with Figure \ref{Fig:ReddeningCoefficients}, the observed distribution of extinction curve parameters lies off the relations predicted by  \citet{1989ApJ...345..245C},   \citet{1999PASP..111...63F}, and \citet{2007ApJ...663..320F}.

That these two ratios have uncorrelated variations disproves the canonical expectation that variations in the shape of the optical+near-IR extinction curve can be explained by a single parameter, $R_{V}$. There are at least two independent degrees of freedom in the optical+near-IR wavelength regime, and the fact that the maximum we can possibly measure with four photometric bandpasses is three degrees of freedom, suggests that there may be more. 

\begin{figure}
\begin{center}
\includegraphics[totalheight=0.25\textheight]{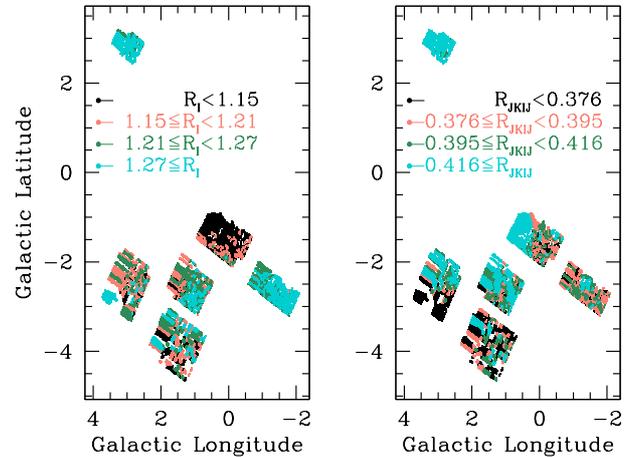}
\end{center}
\caption{\large Spatial distribution as a function of Galactic coordinates (degrees) of the optical extinction coefficient $R_{I}=A_{I}/E(V-I)$ (left panel) and the near-IR extinction coefficient $R_{JKIJ}=E(J-K_{s})/E(I-J)$ (right panel). The colours denote quartiles weighted to have approximately equal surface area. Though both the optical and near-IR extinction coefficients have significant variations across the sky, these variations are uncorrelated.} 
\label{Fig:MapperExtinction}
\end{figure}

In Figure \ref{Fig:MapperExtinction}, we show the distributions of $E(J-K_{s})/E(V-I)$ and $A_{I}/E(V-I)$ as a function of direction. In both cases, adjacent sightlines tend to have similar values of the extinction coefficient, which robustly suggests that the measurements and their variations are significant. The distinct distributions in the left and right panels clearly demonstrate that the variations are largely uncorrelated.

\begin{figure}
\begin{center}
\includegraphics[totalheight=0.17\textheight]{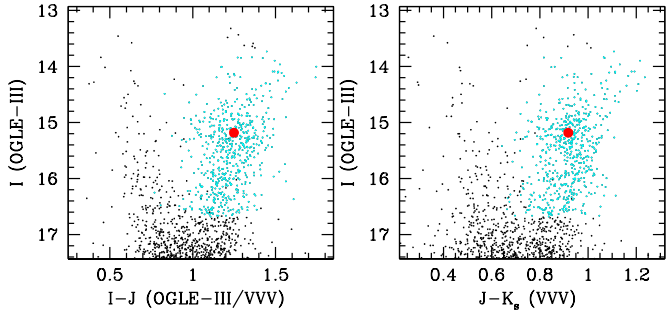}
\end{center}
\caption{\large Colour-magnitude diagrams in $I$ vs $(I-J,J-K_{s})$ toward $(l,b)=(1.68^{\circ},-3.58^{\circ})$.  Symbols are as in Figure \ref{Fig:MarkovPresentationPlot2}. This sightline, with an anomalous large extinction coefficient of $E(J-K_{s})/E(I-J)=0.66$, gives no indication from its CMD of being the product of a systematic such as differential reddening.} 
\label{Fig:MarkovPresentationPlot3}
\end{figure}

Of interest in Figure \ref{Fig:ReddeningCoefficients2}  is a sparse cloud of outliers with much higher values of $E(J-K_{s})/E(I-J)$. These points appear spurious at first glance, but they turn out to be legitimate. In Figure \ref{Fig:MarkovPresentationPlot3}, we show the CMDs for a sightline toward $(l,b)=(1.68^{\circ},-3.58^{\circ})$, with measured extinction coefficient of $E(J-K_{s})/E(I-J)=0.66$, vastly higher than the sample mean of $E(J-K_{s})/E(I-J)=0.40$. The CMDs reveal that the sightline looks fine, there's no confounding issue such as neglected removal of globular cluster contamination, differential reddening, or failed colour-selection. We also verify the photometry by comparing the VISTA photometry to the 2MASS photometry for some of the brighter points, to rule out any potential issues with calibration or observational factors such as the passage of small clouds or bright solar system bodies during the VISTA observations. In both $J$ and $K_{s}$, the differences are usually less than 0.10 mag, and thus the measurements are deemed reliable. 

 Thus, we have to conclude that the story of extinction curve variations toward the inner Milky Way is a much deeper story than that of $R_{V}$ variations.  Further, the cloud of spurious-looking outliers near the top of Figure \ref{Fig:ReddeningCoefficients2} is in fact physically significant. 

\subsection{A Principal Component Analysis of Extinction Coefficient Variations}
Principal component analysis (PCA) is a statistical tool to analyze the number of degrees of freedom of a dataset without the possible bias of needing physical interpretation of the meaning of each variable, thus allowing the data to speak for itself. A set of orthogonal basis vectors is computed by rotating the coordinate-axes in which the variables measured in an n-dimensional space (where ``n" is the number of variables) such that the new variables are uncorrelated (see \citealt{2012AcA....62..269A} for an astronomy-relevant application).  

We compute the principal components over three variables, $0.8352{\times}(A_{I}/E(V-I)$-1.1973), $1.3406{\times}(E(I-J)/E(V-I)-0.7459)$, and $3.3822{\times}(E(J-K_{s})/E(V-I)-0.2965)$. The coefficients \{0.8352, 1.3406, 3.3822\} are chosen such that each input dimension has the same mean value, otherwise the first principal component will be nearly parallel to the largest vector, whereas we are interested in diagnosing extinction curve variations consistently over the entire wavelength regime. Principal component decomposition automatically subtracts the means of the three vectors: \{1.1973,0.7459,0.2965\}. 

\begin{figure}
\begin{center}
\includegraphics[totalheight=0.34\textheight]{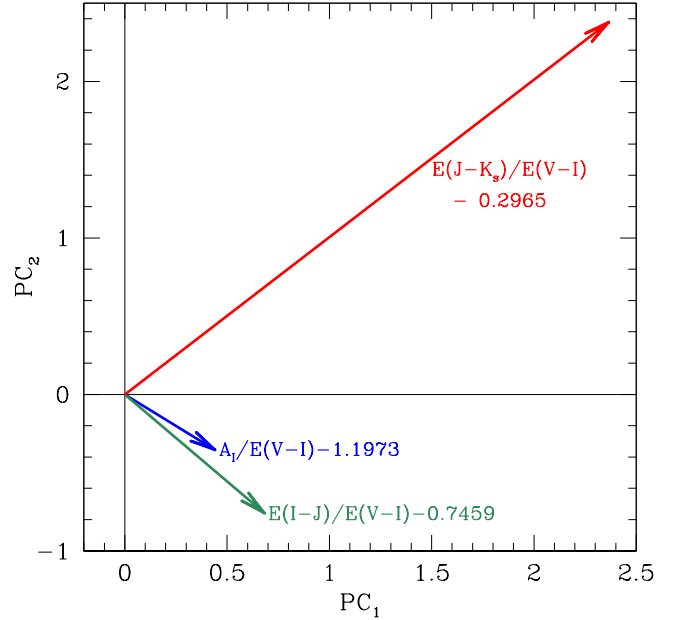}
\end{center}
\caption{\large Projection of the three independent extinction ratios onto the plane of the first two principal components. In this parameterization,  the lengths of the vectors correspond to the inverse as their average value in our data set. These two principal components describe 73\% and 22\% of the variance of the three extinction ratios.  } 
\label{Fig:PrincipalComponent}
\end{figure}

The three eigenvalues of the principal component de-composition are 0.0117, 0.0034, and 0.0008, corresponding to standard deviations along the axes of $\sim$(11, 6, 3)\% in the three rotated reddening ratios. The three principal components derived contribute 73\%, 22\%, and 5\% of the total variance, consistent with the claim made in the prior section that we find two degrees of freedom to the extinction curve in our dataset. The projection of the reddening vectors onto the principal component space is shown in Figure \ref{Fig:PrincipalComponent}. The first two principal components are equal to:
\begin{equation}
\begin{split}
PC_{1} = 0.6340 (A_{I}/E(V-I) - 1.1973)\\ 
+ 0.3555(E(I-J)/E(V-I)-0.7459) \\
+ 0.2081 (E(J-K_{s})/E(V-I)-0.2965)
\end{split}
\end{equation}
\begin{equation}
\begin{split}
PC_{2} = -0.5088 (A_{I}/E(V-I)- 1.1973)  \\
-0.4229( E(I-J)/E(V-I)-0.7459) \\
+ 0.2092 (E(J-K_{s})/E(V-I)-0.2965)
\end{split}
\end{equation}

\subsection{Critical Boundary Value: Relative Extinction in $V$ and $K_{s}$ }
We can estimate extinction in different filters with a conversion such as the following:
\begin{equation}
A_{Ks} = A_{I} - E(I-J) - E(J-K_{s}).
\label{EQ:AKS}
\end{equation}
Though Equation \ref{EQ:AKS} has the advantage of being analytically exact, it has the disadvantage of producing extinction measurements with correlated errors, as the error in $A_{I}$ enters linearly into $A_{Ks}$, which is why the majority of the analysis in this paper focuses on the (independent) measurements of the colour excesses. 

Regardless, the ratios should still be reliable in the median, for which we measure:
\begin{equation}
\mid \frac {A_{V}} {A_{Ks}} \mid = 13.44.
\end{equation}
That is a considerably greater ratio than the canonical value of 8.25  \citep{1989ApJ...345..245C}. The three median ratios measured in this work, $A_{V}:A_{I}:A_{J}:A_{Ks}$, are $1:1.85:4.84:13.44$. We report the median rather than the mean as the mean is distorted by a small number of sightlines with considerable errors, leading to unphysically small values of $A_{Ks}$.

\subsection{Construction of Better Wesenheit Functions to Minimize the Effects of Extinction}
\label{sec:UberWesenheit}
In various fields of astronomy such as the cosmological distance ladder, \textit{Wesenheit}\footnote{``Wesenheit" is the German word for ``essence" or ``nature".} functions are used to minimize the dependence of extinction on apparent magnitudes and thus distances \citep{1982ApJ...253..575M,2011ApJ...741L..36M,2011ApJ...733..124S,2015MNRAS.451..724W}. This is done by subtracting from the apparent magnitude a colour term where the slope is believed to be the average total-to-selective extinction ratio, for example:
\begin{equation}
W_{I} = I - 1.45(V-I),
\label{EQ:Wesenheit}
\end{equation}
is commonly used, and has some empirical support toward sightlines such as the LMC \citep{2003ApJ...590..284U,2012ApJ...748..107P}. Though Equation \ref{EQ:Wesenheit} no doubt performs very well over large swaths of the sky, we have demonstrated in this work that it fails spectacularly toward the inner Milky Way. We have also demonstrated that there is no single universal extinction curve for this Galaxy, and thus it is safe to assume that the same applies to other galaxies. Thus, such simple Wesenheit functions should usually be done away with in this era of precision cosmology. 

An alternative, as per the fits seen in Figure \ref{Fig:ReddeningCoefficients}, the use of \textit{uber\footnote{``uber" is the German word for ``above" or ``at a higher level".}-Wesenheit} functions, such that the apparent magnitude is insensitive to not only variations in extinction assuming a mean extinction curve, but also the dominant first-order variations in the extinction curve. We remove sources with high differential reddening ($\sigma_{(V-I)RC} \geq 0.18$), poor fits ($\sigma_{RC} \geq 0.30$), very high-reddening values that increase the odds of potential systematics such as incompleteness ($(V-I)_{RC} \geq 4.30$) and extreme values of the extreme coefficients ($A_{I}/E(V-I) \geq 1.45$, $E(J-K_{s})/E(I-J) \geq 0.46$). We recursively remove 3-$\sigma$ outliers and obtain the following relation on the $VIJ$ plane:
\begin{equation}
A_{I} = 0.1333E(V-I)+1.4254E(I-J)
\label{EQ:RIJVI}
\end{equation}
It is a tighter relation, with a correlation coefficient $\rho=0.8194$, and can also be discerned from Figure \ref{Fig:ReddeningCoefficients}. A serendipitous result emerges: the coefficient of $E(V-I)$ is very small, only 9\% the size of the coefficient of $E(I-J)$. In practice, it turns out than that the total-to-selective extinction ratio $A_{I}/E(I-J)$ has very little dependence on extinction curve variations. The mean value is given by $<A_{I}/E(I-J)>=1.6063$, and the 1-$\sigma$ scatter by $\sigma_{A_{I}/E(I-J)}=0.066$, or 4.1\%. In contrast, the scatter we expect just from the statistical measurement error in $I_{RC}$ is 2.3\%, and thus the intrinsic scatter in $A_{I}/E(I-J)$ is as small as 3.4\% in our sample. 

We thus suggest:
\begin{equation}
W_{I} = I - 1.61(I-J),
\end{equation}
as a surprisingly robust Wesenheit magnitude. We note that the predicted extinction coefficients of $A_{I}/E(I-J)$ from \citet{1989ApJ...345..245C},   \citet{1999PASP..111...63F}, and \citet{2007ApJ...663..320F} are 1.93, 1.89, and 1.79 respectively. 

\section{Comparisons to Prior Investigations}
\label{sec:Comparison}

\subsection{Shorter-Wavelength Photometry}
In principle it would be interesting to map extinction curve variations over the broadest possible wavelength, which should become possible over time as more photometry of the Galactic bulge is taken. 

One study available for comparison is that of \citet{2010A&A...515A..49R}, who measured photometry of the ``Chandra bulge field" (toward $(l,b)=(0.11^{\circ},-1.43^{\circ})$) in a diverse array of filters with HST's \textit{Advanced Camera for Surveys} (ACS). Unfortunately, we cannot make a direct comparison as they did not publish their input data, only their final results. They report $A_{F625W,\rm{ACS}}/(A_{F435W,\rm{ACS}}-A_{F625W,\rm{ACS}}) =1.25 \pm 0.09$. Their error was the uncertainty on the regression, which emerges due to both measurement errors and the genuine variations in the underlying extinction curve. Given the latter source of error, their quoted error is actually likely to be an overestimate. The predicted coefficients from standard extinction curves of \citet{1989ApJ...345..245C},   \citet{1999PASP..111...63F}, and \citet{2007ApJ...663..320F} are approximately 1.92, 1.64, and 1.64 respectively. \citet{2010A&A...515A..49R} thus measured a steeper-than-standard extinction curve toward those sightlines, regardless of how one defines ``standard". The predicted extinction curves are either $R_{V}=1.97$ \citep{1989ApJ...345..245C} or  $R_{V}=2.46$ \citep{1999PASP..111...63F}.

The typical extinction coefficients we measure toward those sightlines are $A_{I}/E(V-I)=1.10$, $E(I-J)/E(V-I)=0.70$, and $E(J-K_{s})/E(V-I)=0.25$. Interestingly, we cannot find an $R_{V}$ match even if we restrict the fit to the optical filters. Fitting $A_{I}/E(V-I)=1.10$ requires $R_{V} \approx 2.20$ in either the parameterization of \citet{1989ApJ...345..245C} or that of  \citet{1999PASP..111...63F}. The resulting predicted values of $A_{F625W,\rm{ACS}}/(A_{F435W,\rm{ACS}}-A_{F625W,\rm{ACS}})$ are 1.39 and 1.09 respectively, both failing to match the result of  \citet{2010A&A...515A..49R}, with equal errors of opposite signs. We list all of the implied values of $R_{V}$ toward this sightline in Table \ref{table:Revnivtsev}. 

\begin{table}
\centering
\begin{tabular}{|l|l|ll|}
	\hline \hline
Index & Value & $R_{V,\rm{C89}}$ & $R_{V,\rm{F99}}$  \\
	\hline \hline \hline
$A_{F625W}/(A_{F435W}-A_{F625W})$ & 1.25 & 1.97 & 2.46 \\
$A_{I}/E(V-I)$ & 1.10 & 2.20 & 2.21 \\
$E(I-J)/E(V-I)$ & 0.70 & 2.72 & 3.07 \\
$E(J-K_{s})/E(V-I)$ & $0.25$ & 1.97 & 0.98 \\
$E(J-K_{s})/E(I-J)$ & $0.36$ & 2.69 & 3.67 \\
	\hline
\end{tabular}
\caption{The best-fit values of $R_{V}$ as a function of extinction curve parameter for the parameterizations of \citet{1989ApJ...345..245C} and \citet{1999PASP..111...63F} for the sightline investigated by \citet{2010A&A...515A..49R}. }
\label{table:Revnivtsev}
\end{table}
\vspace{-0.00in}

\subsection{Measurements of Diffuse Interstellar Bands}
\label{sec:Comparison2}
The correlation between the diffuse intstellar band located at $\lambda_{0} = 15272.42$\,\AA\, and interstellar reddening was measured by \citet{2015ApJ...798...35Z}:
\begin{equation}
<W_{DIB}> = 12.2572*E(H-[4.5 \mu]),
\end{equation}
where $W_{DIB}$ is the diffuse interstellar band equivalent width in milli-angstroms, and the reddening $E(H-[4.5 \mu ])$ is taken from \citet{2012ApJS..201...35N}. The function reported by \citet{2015ApJ...798...35Z} is in terms of $A_{V}$, which was extracted from the measurements of \citet{2011ApJ...739...25M} with the conversion factors $A_{V}= 8.8A_{Ks}$, $A_{Ks} = 0.918E(H-[4.5 \mu])$.

We match our catalogue of reddening and extinction curve variations with that of of $W_{DIB}$ and $E(H-[4.5 \mu])$ measurements satisfying   ($|l| \leq 5^{\circ}$, $|b| \leq 5^{\circ}$)   from \citet{2015ApJ...798...35Z}. We obtain a paltry 23 matches, due to poor spatial overlap. In order to expand the sample, we match the DIB catalogue of \citet{2015ApJ...798...35Z} to the $E(J-K_{s})/E(V-I)$ catalogue of \citet{2013ApJ...769...88N}, which is the extinction ratio most reliably measured in that work. This yields 137 matches.

\begin{figure}
\begin{center}
\includegraphics[totalheight=0.35\textheight]{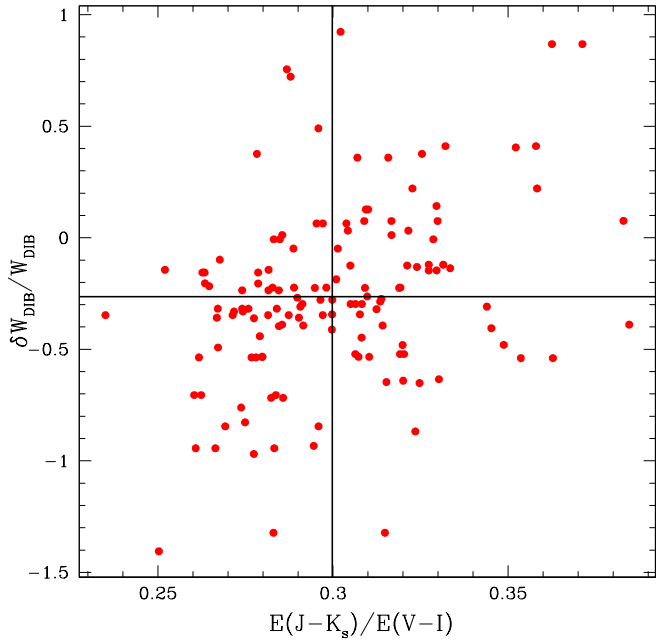}
\end{center}
\caption{\large Scatter of residual of diffuse interstellar band strength relative to predictions of \citet{2015ApJ...798...35Z}, versus the $E(J-K_{s})/E(V-I)$ measurements from \citet{2013ApJ...769...88N}, shown as red points. There is a slight, positive correlation, $\rho=+0.34$. The thick black lines denote the mean values of the two variables for the 137 data points. The mean 1-$\sigma$ measurement error in ${\delta}W_{DIB}/W_{DIB}$ is 0.27.} 
\label{Fig:DIBcomparison}
\end{figure}

Then, from each $W_{DIB}$ measurement, we subtract the predicted measurement to obtain a residual:
\begin{equation}
\frac{{\delta}W_{DIB}}{W_{DIB}} = \frac{W_{DIB} - 12.2572*E(H-[4.5 \mu])}{W_{DIB}}.
\end{equation}
One might expect the residuals to be randomly distributed and have a mean of zero. However, we instead find a correlation of $\rho=+0.34$ between ${\delta}W_{DIB}/W_{DIB}$  and $E(J-K_{s})/E(V-I)$. We show the scatter in Figure \ref{Fig:DIBcomparison}. The p-value for the correlation is $4.6 {\times}10^{-5}$ -- the odds of deriving this correlation by chance are $\sim$21,000:1. 

Interestingly, the mean value in ${\delta}W_{DIB}/W_{DIB}$    is $-0.26$. This is extremely unlikely to be due to chance as the scatter measured by \citet{2015ApJ...798...35Z} was $\sim$50\% per star, and thus our sample mean is a $\sim$6.2$\sigma$ outlier. This offset is consistent with the accumulating evidence that the interstellar medium toward the inner Milky Way has systematically different properties to that elsewhere in the Galaxy. The correlation between  $E(J-K_{s})/E(V-I)$ and ${\delta}W_{DIB}/W_{DIB}$ implies a ``standard" value of ${\delta}W_{DIB}/W_{DIB}$ would be reached in the mean if a ``standard" value of $E(J-K_{s})/E(V-I)$ is also reached in the mean. 

This suggests that the ratio between diffuse interstellar band strength and interstellar extinction may depend on the properties of interstellar medium. This is not surprising, given observations of sightline dependence for other diffuse interstellar bands \citep{2013ApJ...774...72K}, though it is the first demonstration for the $\lambda_{0} = 15272.42$\,\AA\,  diffuse interstellar band. This issue warrants further investigation. We point to the recent measurement of five distinct diffuse interstellar band equivalent widths toward the type Ia supernova 2014J by \citet{2015MNRAS.451.4104J}, which is located behind anomalous dust \citep{2014ApJ...784L..12G}, as an example of potential future applications. 

\subsection{The Planck Reddening Maps}
\begin{figure}
\begin{center}
\includegraphics[totalheight=0.33\textheight]{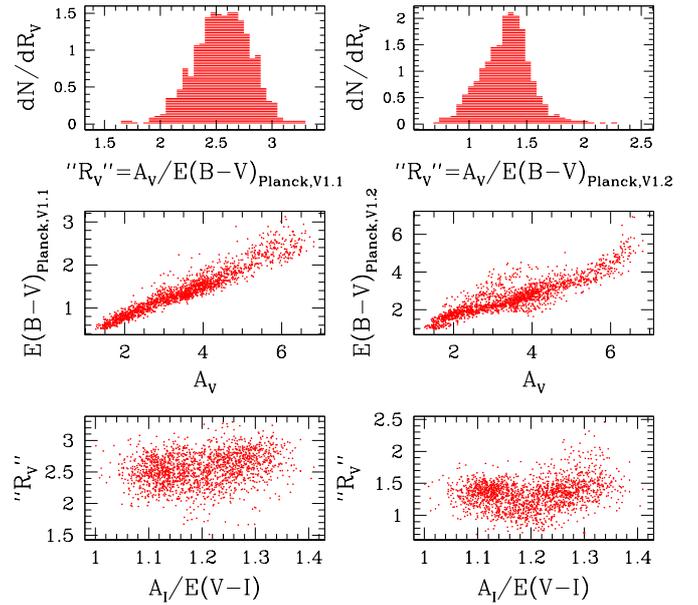}
\end{center}
\caption{\large \textit{Planck} determinations of $E(B-V)$ from the version 1.1 maps (left panels) and 1.2 maps (right panels) as a function of $A_{V}$ in the top and middle panels, and $A_{I}/E(B-V)$ in the bottom panels.  The version 1.1 maps do better in the mean than the version 1.2 maps. Neither version appears sensitive to extinction curve variations. } 
\label{Fig:PlanckHist}
\end{figure}

We compare our reddening measurements to the version 1.1 $E(B-V)$ all-sky maps from \textit{Planck}, and the version 1.2 maps  \citep{2014A&A...571A..11P}.  The \textit{Planck} maps report $E(B-V)$, which we can compare to our measurements of $A_{V}$ to obtain a fiducial $R_{V}$. We show various diagnostics in Figure \ref{Fig:PlanckHist}.  We find that neither reddening map works well, both have unexplained scatter, but the bias in the version 1.2 maps relative to the measured reddening reaching a catastrophic and colossal $\sim$100\%. 

The version 1.1 maps (left panels) yield a mean and standard deviation of $R_{V}=2.55 \pm 0.25$, both plausible given the other measurements in this work. However, in the bottom panel we see that the suggested $R_{V}$ is uncorrelated with $A_{I}/E(V-I)$ ($\rho=0.25$), with the correlation dropping to $\rho=0.20$ and $\rho=0.03$ for $E(I-J)/E(V-I)$ and $E(J-K_{s})/E(V-I)$ respectively.  In contrast, the correlations between $A_{I}/E(V-I)$ and $E(I-J)/E(V-I)$ and $E(J-K_{s})/E(V-I)$ were $\rho=0.80$ and $\rho=0.57$, respectively, see Figure \ref{Fig:ReddeningCoefficients}.  The fact that all of these correlations are small suggests that extinction curve variations are not responsible for the offset between reddenings inferred from infra-red emission and that measured from stellar colours, and that there is another source of ``error" at play. 

The version 1.2 maps (right panels of Figure \ref{Fig:PlanckHist}) yield a mean and standard deviation of $R_{V}=1.33\pm 0.22$, which is not plausible given the other measurements in this work, and suggests that reddening in the version 2 maps is overestimated by a factor of 2. In the bottom panel we see that the suggested $R_{V}$ is uncorrelated with $A_{I}/E(V-I)$ ($\rho=0.20$), which seems unlikely, though it would be beneficial to obtain $B$-band photometry of the bulge in order to be sure. It also appears to have a strange quadratic behaviour, with $R_{V}$ minimized at $A_{I}/E(V-I) \approx 1.20$,  approximately the mean value in our sample. 

The effect of background extinction, whereby emission from dust located behind the red clump stars, could in principal be a potential bias to the results. However, we expect to be small, as our sightlines typically intersect the bulge at a height $\sim$300 pc above the plane. We measure the Pearson correlation coefficient between the ratio of measured reddening $E(V-I)$ and $E(B-V)$ from the Planck v1.1 maps to absolute latitude (a proxy for separation from the plane) to be $\rho=+0.052$, i.e.:
\begin{equation}
\rho \biggl( \frac{E(V-I)}{E(B-V)_{Planck\,v1.1}} , |b|      \biggl) = +0.052.    
\end{equation}
The Pearson correlation coefficient if we instead use the Planck v1.2 maps is $\rho=-0.147$. If background emission were a significant source of error in the analysis, then the ratio of measured to expected reddening would drop rapidly with decreasing absolute latitude, in other words there would be a strong positive correlation. We do not find a large, positive value of $\rho$ with either map, in agreement with our expectation that the systematic error from background emission is small. 

The non-linearities that \citet{2014MNRAS.445.4252W} identified when comparing the \textit{Planck} reddening maps to photometry of QSOs are not present in our comparison, furthering the argument that they are due to zero point calibrations. Our methodology will necessarily be less sensitive to zero-point calibrations, as the reddening values probed in this work are $\sim10{\times}$ higher than those probed by \citet{2014MNRAS.445.4252W}. What is consistent between our two works are that the version 1.1 map is accurate in the mean whereas the version 1.2 map overestimate reddening by a factor $\sim 2$. This consistency is impressive given the different methodology: \citet{2014MNRAS.445.4252W} used $ugriz$ photometry to study reddening toward QSOs in halo sightlines spanning $\sim$10,000 deg$^{2}$, and thus probed dust predominantly from the solar neighbourhood, with a normalization of $E(B-V) \lesssim 0.20$. 

These discrepancies will ultimately require more resolution, more wavelength coverage and superior comparison with models to resolve. Of possible interest may be the dust model of \citet{2013A&A...558A..62J}, which incorporate different distributions of small carbon grains and larger silicate/iron grains.

\section{Summary}
\label{sec:Summary}
In this investigation we have combined $VIJK_{s}$ photometry from the OGLE-III and $VVV$ surveys to make nearly 2,000 independent measurements of each of $A_{I}$, $E(V-I)$, $E(I-J)$, and $E(J-K_{s})$ toward the bulge. We have done so over a range of coordinates within which metallicity variations are small, and for which distance effects due to the Galactic bar can be accounted for. 

We confirm previous reports that the extinction curve toward the inner Milky Way is variable and non-standard \citep{2003ApJ...590..284U,2009MNRAS.394.2247G,2013ApJ...769...88N}. Furthermore, not only is the extinction curve non-standard in the mean, it is also poorly fit by the parameterizations of \citet{1989ApJ...345..245C} and  \citet{1999PASP..111...63F} regardless of how $R_{V}$ is varied. The mean Galactic extinction curve of \citet{2007ApJ...663..320F} is also a poor fit. These fits are poor both with respect to the mean of the Galactic bulge extinction curve, as well as the fact they never intersect the variations thereof. 

We find that the shape of the interstellar extinction has at least two degrees of freedom, as the variations in $A_{I}/E(V-I)$ and $E(J-K_{s})/E(I-J)$ are uncorrelated. We use principal component analysis to confirm the presence of two significant independent degrees of freedom in our data. This suggests a relatively large, and completely undiagnosed, source of systematic errors in cosmological investigations of Cepheids and Type Ia supernovae. 

We look forward to extending our investigations over a broader range of wavelengths, for example by incorporating photometry from the Dark Energy Camera \citep{2008SPIE.7014E..0ED} and Pan-STAARS \citep{2012ApJ...750...99T}. Further insights may be gleaned by comparison to measurements of diffuse interstellar bands from surveys such as APOGEE (\citealt{2015ApJ...798...35Z}, a comparison already begun in this work) GALAH \citep{2015MNRAS.449.2604D}, and Gaia-ESO \citep{2015A&A...573A..35P}.

\section{Acknowledgments}
We thank the referee for a helpful review of the manuscript.

We thank Andrew Gould, Carine Babusiaux, Albert Zijlstra, and  Edward Schlafly for helpful discussions.

D.M.N was primarily supported by the Australian Research Council grant FL110100012. This research was supported in part by the National Science Foundation under Grant No. NSF PHY11-25915. 

The OGLE project has received funding from the National Science Centre, Poland, grant MAESTRO 2014/14/A/ST9/00121 to AU.

We gratefully acknowledge the use of data from the ESO Public Survey program ID 179.B-2002 taken with the VISTA telescope, data products from the Cambridge Astronomical Survey Unit. Support for the authors is provided by the BASAL CATA Center for Astrophysics and Associated Technologies through grant PFB-06, and the Ministry for the Economy, Development, and Tourism's Programa Iniciativa Cientifica Milenio through grant IC120009, awarded to Millenium Institute of Astrophysics (MAS). D.M. and M.Z. acknowledge support from FONDECYT Regular grant No. 1130196 and 1150345, respectively. R.K.S. acknowledges support from CNPq/Brazil through projects 310636/2013-2 and 481468/2013-7. J. A.-G. acknowledges support from Fondecyt Postdoctoral project 3130552 and FIC-R Fund project 30321072

This publication makes use of data products from the Two Micron All Sky Survey, which is a joint project of the University of Massachusetts and the Infrared Processing and Analysis Center/California Institute of Technology, funded by the National Aeronautics and Space Administration and the National Science Foundation.

\bibliography{Reddening_VVV_OGLE_MNRASv2}

\newpage

\appendix

\end{document}